Na prawach rękopisu

Instytut Informatyki
Politechnika Wrocławska# GPU-Based Massive Parallel Kawasaki Kinetics In Monte Carlo Modelling of Lipid Microdomains

Seria PRE nr 5Mateusz Lis
Łukasz PintalSłowa kluczowe:
  Biomembrane
  Monte Carlo simulation
  Lipids
  GPU-computingKrótkie streszczenie:
This paper introduces novel method of simulation of lipid biomembranes based on1

Metropolis-Hastings algorithm and Graphic Processing Unit computational power. Method gives up to 55 times computational boost in comparison to classical computations. Extensive study of algorithm correctness is provided. Analysis of simulation results and results obtained with classical simulation methodologies are presented.

# Introduction

Problem of lipid microdomains creation is currently under extensive computational (as well as experimental) studies (Petra Schwile and others). Computational studies of such systems often suffer due to limited computational resources and simulation size scales irrelevant in comparison to experimental. In recent years, massive parallel computing has become a valuable tool in computational studies (Ising models).

It is very hard to get real statistics of clusters sizes in the lattice of 400x400 for larger Gibbs energies between lipids, and statistics of small cluster (nanoclusters) sizes for small Gibbs energies. However, computational power required to perform larger scale simulation may constrain modern researches. In this paper we provide a class of algorithms which help to overcome this problem. Basic idea is to utilize power of parallel execution using Graphic cards multiprocessors. Various studies (source) often overcome this problem by application of different transition functions. However, to study dynamics of lipids domains, Kawasaki Kinetics [10] give better description of time evolution of the investigated systems, especially when it comes to lipid microdomains. A number of various studies presented approaches of parallel computing to Ising models, but hardly any of them touched the problem of Kawasaki kinetics.

Schulz et al. (2009) [3] reported 24 times increase in speed of calculations when applied their domain decomposition algorithm to integration of KPZ equation using Kawasaki kinetics based on GPU.

# Materials and Methods

### GPU architecture

There are many written sources about GPU (Graphic Processing Units) Architecture [1] [5]. The goal of this section, however, is to provide understanding of basic concepts that were used to design simulation algorithms in this paper. For more extensive information on GPU architecture, please refer to mentioned resources.

CPU (Central Processing Unit) of modern computer is designed to provide vast amount of flexible computational resources to a single task. In case of large data processing, we often reach limits of sequential calculations. In cases, where spatial decomposition of the problem is possible, instead of sequential processing, a parallel approach can be applied. In sequential computations, it is assumed that single tasks have to be executed one after another. Parallel approach creates possibility of simultaneous execution of computational tasks. Recent development in GPGPU (General Purpose Graphic Processing Unit) allowed to utilize massive parallelization in the order of thousands of processing units localized on GPU, whereas in case of CPU parallelization scale is in the order of tens. Usual small computational power and flexibility of each GPU processing unit is compensated by massive parallel execution. This is illustrated in the [Fig. 1], where in the CPU part there are large Control and Cache units which provide flexibility in the instruction set design, whereas GPU consists of vast amount of small blocks of processing units and their control units which are able to perform simple computational instructions simultaneously on large data sets.

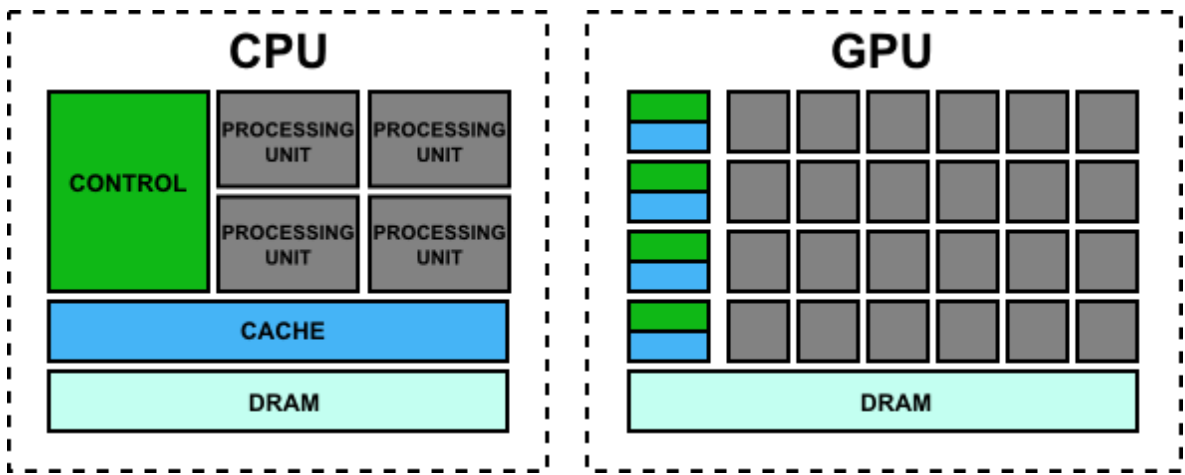

Fig. 1. Comparison between CPU and GPU architecture.

The advantage of GPU processing is a field of fierce discussion and multiple authors provide different results when it comes to comparison between CPU and GPU performance [7]. The scale of computational speedup varies between 1.5 [1] and 100 or even more [8] and is a matter of applied problem an optimizations. What is commonly agreed, is that GPU processing provides substantial boost for problems that involve simple single precision calculations on large data sets that do not require synchronization [5]. Because of that, rapid development of GPGPU software libraries in recent years can be observed. This enables wide range of applications of this technology. Different technology solutions (CUDA, DirectCompute, OpenCL) serve various applications from desktop workstations to computational clusters.

Much of the challenge in design of efficient GPU applications lies in proper decomposition of the problem to available computational resources. Every computational thread in GPU is capable of performing simple tasks on specified part of data. A group of threads is called block, which size varies on different hardware devices (up to 1024 threads per block). The efficiency of GPU device depends on the number of computational units (cores) that are able to handle operations performed in a single thread. Modern GPU architecture provides up to 512 cores on a single device (nVidia Tesla M2090) [6]. Threads in one block can share fast access memory called *shared memory.* Memory shared between the blocks is much slower. Thus, exchanging data between blocks is to be avoided, as well as threads synchronization which makes parallel close to sequential processing. Another important issue is moving data between computer and device memory, which is compulsory, but should be reduced to minimum, as it is time-consuming. In perfect situation, it is possible to copy whole data to device memory, then perform simulations and move it back after computations. However, saving intermediate results usually requires computer access to the data (it especially affects simulations studies, where complete trajectory of the system is needed to be exported during the computations). Therefore, a successful massive parallel algorithm implementation needs to deal with three major design issues:
  - Data decomposition between blocks and threads.
  - Operations synchronization.
  - Memory management.

Each of these issues will be addressed in the algorithm description section.

**Monte Carlo model with Kawasaki based kinetics**

Goal of the Metropolis algorithm is to generate samples from a particular probability distribution which exists in reality. In case of biological membranes, we would like to sample from possible lipid configurations on membrane's surface. To achieve that goal, we use conditional probability distribution which is based on the state we currently occupy $q(x^*/x(n))$ (where $x^*$ is the candidate state, and $x(n)$ is the state we are currently in). Iteratively, we accept or reject generated candidate samples. Decision is based upon criterion introduced by Metropolis [9]. The Metropolis algorithm proceeds as follows:
1. Generate initial state $x(0)$.
2. For i=1 to N:
    a. Generate candidate state $x^*$ according to q distribution.
    b. Sample $u \sim U[0,1]$
    c. if $u < \min\{ 1, p(x^*)/p(x(i-1)* q(x(i)|x^*)/q(x^*|x(i)) \}$ $x(i) = x^*$, else $x(i) = x(i-1)$

To ensure proper results of simulation, transition distribution $q(x^*|x(n))$ has to satisfy two conditions:
  - ergodicity,
  - detailed balance.

Standard Metropolis algorithm has naturally sequential form, therefore it is hard to parallelize [11]. However, as we show in the next section, for special cases with appropriate transition distribution, there is a possibility of parallelization.

Let us consider a triangular lattice model of lipid bilayer surface [Fig. 2]. Each of the sites contains one lipid. To simplify the description, we assume that lattice contains a binary mixture of lipids, therefore every site of the lattice either contains a lipid of type A or B. We use Metropolis-Hastings algorithm to simulate evolution of the system. Single sample in the Metropolis algorithm denotes state of whole lattice (thus, the candidate probability distribution has as many dimensions as the size of the lattice). We assume there is no lipid exchange with the environment, thus the number of lipids of each type is constant throughout whole simulation. Because of that, we utilize Kawasaki kinetics [10] as a transition distribution for lattices, which ensures constant number of lipids of each type. More recent papers report other transition distribution, which is depicted on the [Fig. 3.] . The idea is to exchange lipids not only from their neighbourhood, but from the whole lattice. This way is reported to require much less steps to achieve equilibration, nevertheless its dynamics are unnatural, therefore it is not applicable to simulations focused on dynamics of the system or non-equilibrium properties. Thus, in this paper we consider Kawasaki dynamics as a reference transition distribution.

To calculate $\min\{1, p(x^*)/p(x(i-1)) * q(x(i-1)|x^*)/q(x^*|x(i-1))\}$ we employ model presented by Almeida in [13]. We assume, that lipid interactions are fully defined using the interaction energy between two particular lipids A and B:
$\omega_{AB} = g_{AB} - \frac{1}{2}(g_{AA} + g_{BB})$
Where $g_{AA}$, $g_{AB}$ and $g_{BB}$ are Gibbs free energies of interaction between A and B. Thus, whole complex system has only one parameter. Nevertheless, multiple [13][15] researches report that this model is capable of providing biologically relevant results, especially in the field of latent membrane organization, or microdomain formation.

As the computational power increases in time, research studies report application of models of similar computational complexity [12][13][14] on growing lattice sizes 50x50 (~1995), 100x100 (~2005), 400x400 (after 2010) lipids. As we show in the results section, those studies may still suffer from errors caused by too small size of lattice, especially when considering lipid microdomains size distribution. Next section presents an alternative transition distribution, which not only enables parallel computations power to be applied to Monte Carlo simulation, but also preserves Kawasaki-like dynamics of the system.

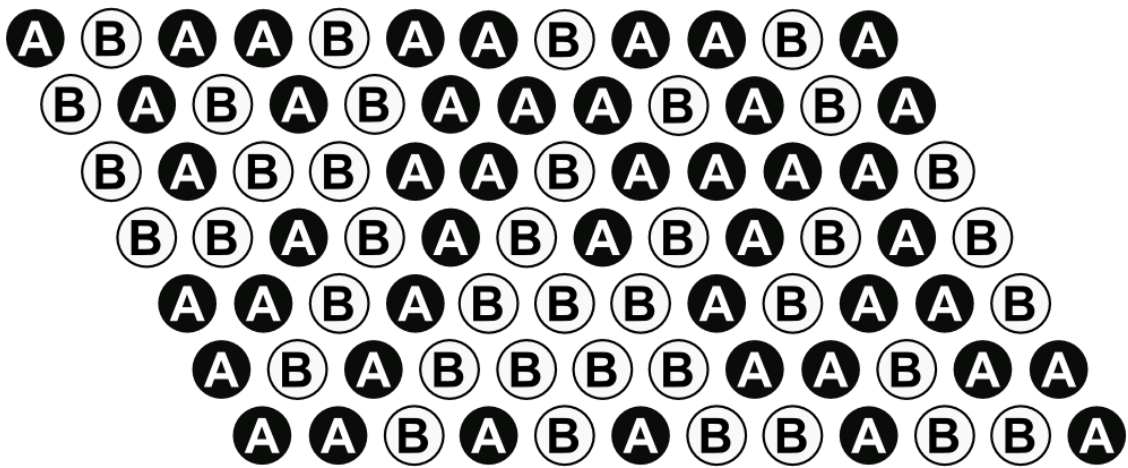

Fig. 2. Triangular lattice model of lipid bilayer surface with randomly distributed lipid sites.

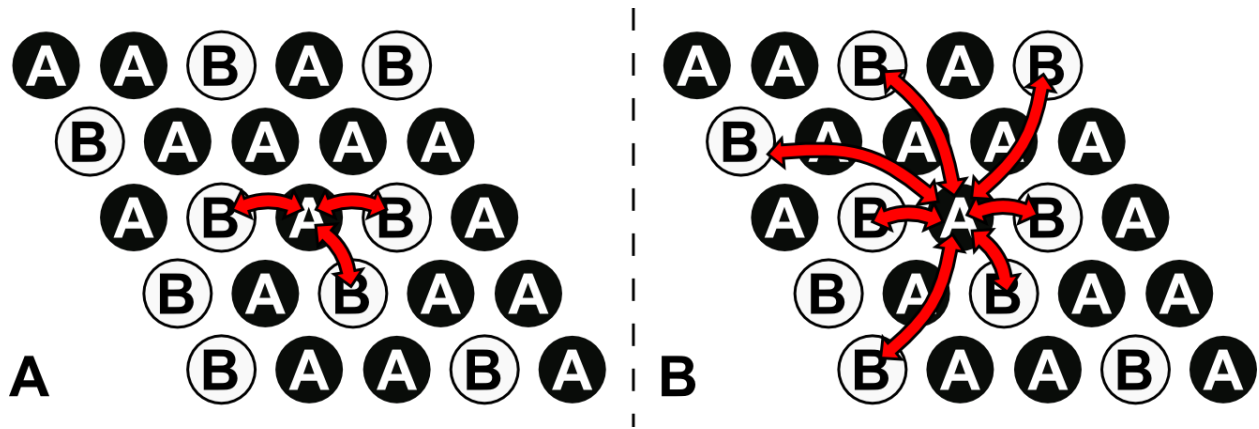

Fig. 3. Comparison of transition distributions: exchanging only direct neighbors (A) and exchanging with all available lattice sites of different types (B).

**Massive Parallel Kawasaki Kinetics**

The idea of introducing massive parallel computations to Kawasaki kinetics was previously considered by [3]. The concept of parallelization was based on spatial lattice decomposition to disjoint domains. Simulations were performed independently for each domain and after a fixed number of iterations, domains were randomly translated. To prevent concurrent memory access, domains had to be separated by small buffers of lipids not taking part in simulations. This approach is reported to be up to 30 times faster than classical Kawasaki method, but it introduces errors due to the independence of simulations in domains and existence of lipids not involved in the simulation process. Those errors depend on the size of single domain and the number of iterations before domains' recalculation. As a result, it introduces a tradeoff between performance and accuracy.

Massive Parallel Kawasaki Kinetics (MPKK) is based on extreme decomposition of triangular lattice to minimal domains that allow Kawasaki transition, namely seven-lipid domains [Fig. 4.A]. It can be shown, that for certain sizes of the lattice, a set of such domains, covering whole lattice exists [Fig. 4. B]. Moreover, if a lattice can be covered in that way, there are exactly seven domain decompositions (coverages) for this lattice. Each of the possible decompositions can be generated using one of the first seven lipids on the lattice as a center point of a starting domain. Thus, a particular decomposition depends only on one parameter, namely the index of the center lipid of the generating domain.

In every step of MPKK simulation a random minimal domain decomposition is generated and a Kawasaki step is performed on each domain simultaneously. Thus, a negative influence of borders and independence of domains is avoided while a high parallelization level can preserved. Due to a large number of domains, a GPU-based parallelization can be utilized, by performing simulation on each domain on a different processing unit. This approach requires domain recalculation in every step. Nevertheless, recalculation is only a matter of generating random integer between 1 and 7.

The MPKK algorithm has the following form:
 1. For i = 1 to N:

a. For j = 1 to 7:
      i. k = rand%7
      ii. generate a decomposition D(k) over k-th lipid
      iii. For each domain d in D(k):
         1. Perform a Kawasaki step on center lipid of d

where N is the number of simulation steps to perform.

Each step of MPKK algorithm requires seven iterations (domain recalculations), to provide compatibility with traditional Kawasaki method, where number of iterations in one step equals the size of the lattice. Comparison between single MPKK and traditional Kawasaki method iteration is depicted on [Fig. 5.].

Properly defined transition distribution in Metropolis algorithm must satisfy certain criteria [Barkema]. First of all, the condition of ergodicity requires transition distribution to enable to reach any state of the system from any other state. Consider single iteration of MPKK. There is non-zero probability of rejection of all proposed lipid exchanges but one. In that case, MPKK performs single Kawasaki exchange in this iteration. Kawasaki kinetics satisfy ergodicity: performing set of those exchanges can move system to any state. Likewise in MPKK, there is non-zero probability to reach any state of lattice by performing set of Kawasaki steps.

Second important requirement is condition of detailed balance defined as follows:
$p_\mu P(\mu \rightarrow \upsilon) = p_\upsilon P(\upsilon \rightarrow \mu)$

Where $\mu$ and $\upsilon$ are possible states of the system. Consider acceptance probability of Metropolis algorithm. Then, to satisfy detailed balance, proposal distribution needs to give equal chance to all possible candidate states. In single iteration of MPKK probability of selecting any particular pair of lipids to exchange is equal to $\frac{1}{7} \cdot \frac{1}{6}$ since every one in seven lipids becomes basis of decomposition and for every lipid there are six neighbors to exchange with. Therefore, proposal distribution is equal for every possible candidate state. Thus, MPKK satisfies detailed balance.

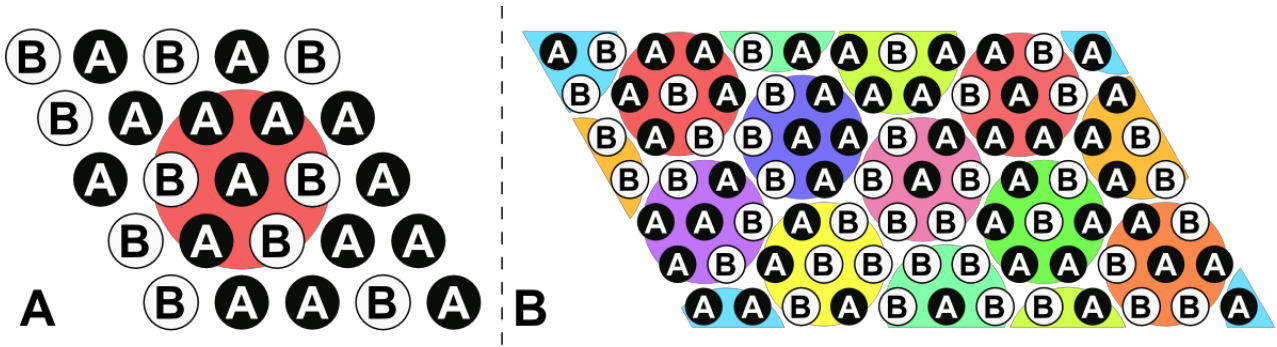

Fig. 4. Minimal (7-lipid) domain (A) and whole lattice coverage by minimal domains (B). Each domain is color coded to depict helical boundary conditions influence.

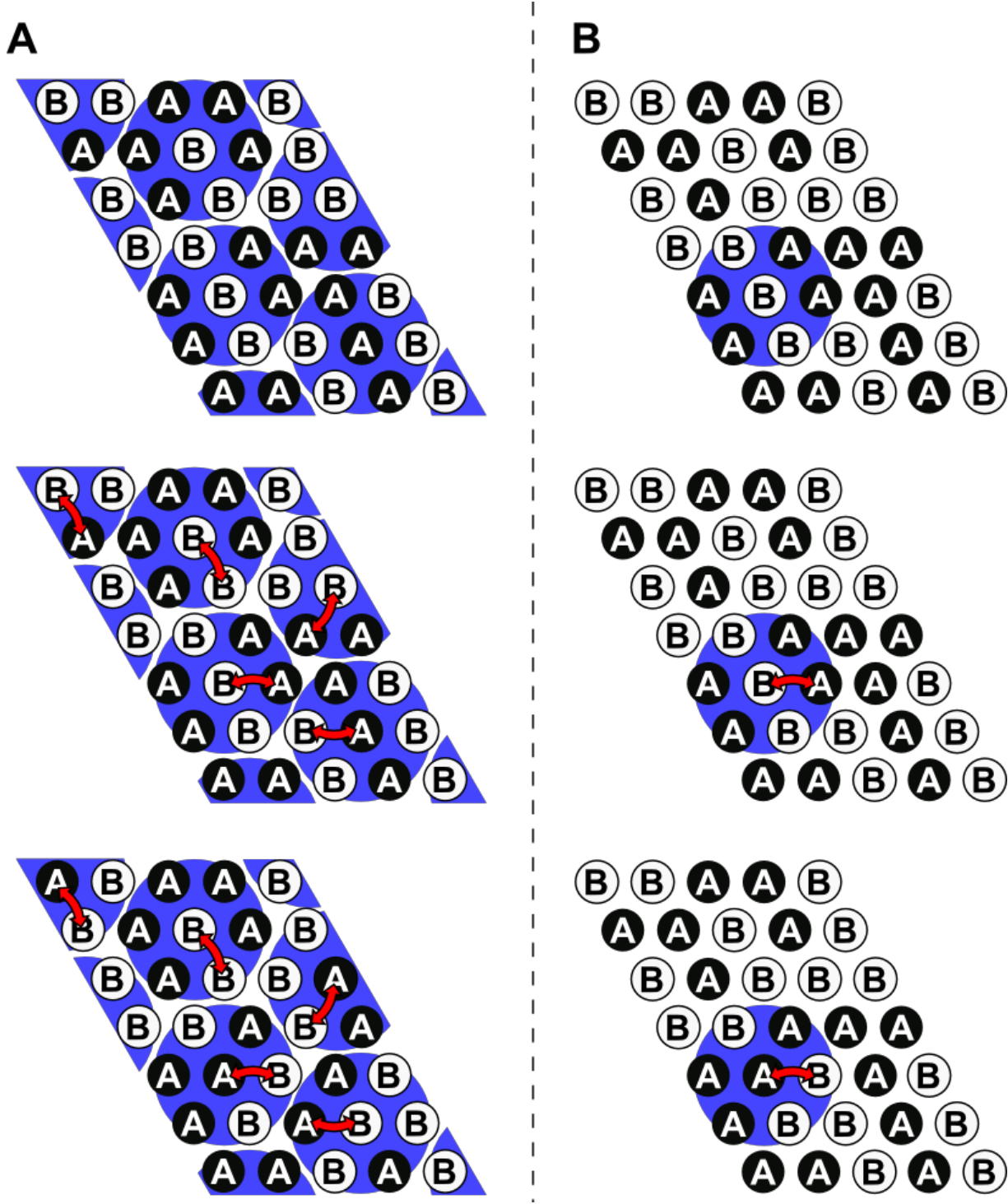

Fig. 5. Comparison between single MPKK (A) and traditional Kawasaki method (B) iteration (please note multiple parallel transitions).

### Image analysis

To facilitate analysis process and reduce disk space required for storing simulation trajectories, snapshots of specific frames were saved in the form of previously scaled images. Each lattice site is depicted as a white or black pixel on image plane, depending on its type. Resulting image is then sloped to preserve original geometry of the triangular lattice and scaled down to desired size. As simulations were performed on GPU device, generation of images was also conducted on the same computational device. This allowed to minimize expensive memory copying operations. Application of supersampling [16] interpolation method enabled preservation of important information stored in trajectory and allowed highly accurate trajectory analysis. It is important to note that performing analytical operations on scaled images is significantly faster than on original data while preserving high quality results. We present fraction of similar first neighbors [17] analysis method as an example analysis method that can be conducted on image trajectory. Example results are presented in results section.

Classical approach for fraction of first neighbors analysis on triangular lattice is the following: for every site in the lattice randomly choose one of its nearest neighbors and increase similar neighbors counter if both sites are of the same type. Result comes from division of similar neighbors counter and lattice size. In case of gray scale images (as supersampling produces gray scale to preserve as much information as possible from the original trajectory), we also randomly choose one of pixel neighbors of every pixel belonging to the lattice. Then to compare pixels we introduce similarity factor - a color space distance between pixels. A simplest method of evaluating similarity of pixels is thresholding. The threshold factor value $\delta_{col} = 35$ has been chosen empirically, but is constant for all performed analysis on different trajectories.

### Cluster Analysis methods

One of the analytical methods presented in the results section is cluster analysis. It enables to assess grouping level of lipids in the domain. In particular, it provides a quantitative method of estimation whether lipids of type A prefer to mix with lipids of type B or not. An efficient cluster calculation method is essential, as cluster analysis needs to be performed regularly to provide partial results and enable averaging over time. Thus, the algorithm complexity of $O(n)$ is required. A Hoshen-Kopelman algorithm [2] was used to satisfy this requirement. A modification applied to the algorithm concerned the triangular lattice structure. In order to adapt Hoshen-Kopelman algorithm, six closest neighbours, calculated using helical boundary conditions, were taken into consideration to assign a label to each lattice site.

### Devices utilized for performance measurements

Performance results presented in results section were calculated on nVidia Tesla M2090 (GPU results) and Intel Xeon CPU E5640 2.67GHz (CPU - 8 cores) with 2.8 GB of memory per core. GPU implementation used CUDA 4.1 library. All implemented algorithms are available in [googlecode/pymd2mc] (provided results used version r104 available at the repository).

# Results

**MPKK vs. Kawasaki method**

Comparison between CPU and GPU simulation results on smaller lattices (up to 100x100) led to a conclusion that MPKK method generates the same results as traditional Kawasaki Method [Fig. 6.].

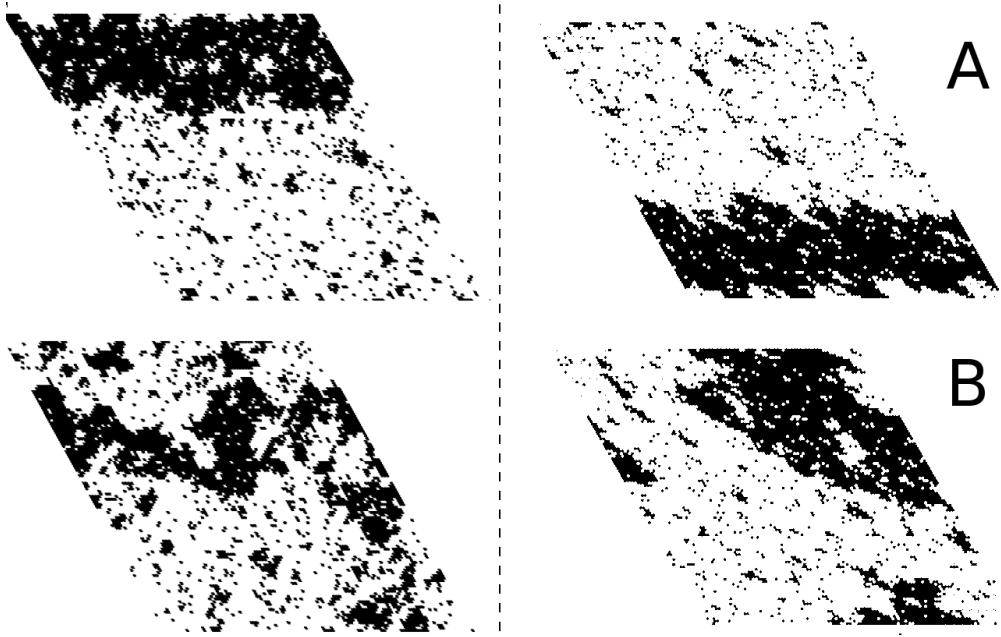

Fig. 6. Comparison of snapshots generated using GPU computations (left side) and calculated on CPU (right). Top half of the image contains snapshot after 10000 of simulation steps whereas bottom snapshots are taken after 10 milion steps. Simulations were started from non-random configuration.

Fraction of first neighbors comparison of discussed methods is depicted in [Fig. 7.].

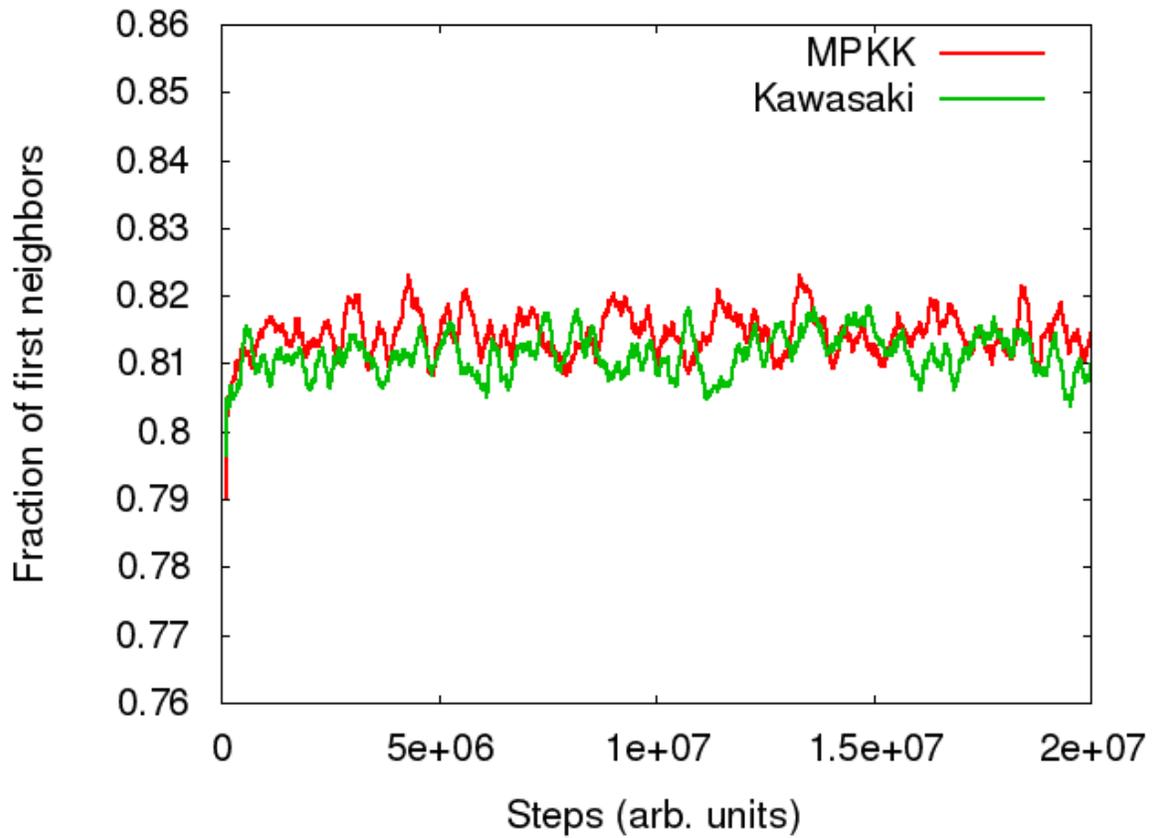

Fig. 7. Comparison between single MPKK (A) and traditional Kawasaki method (B) simulation results: fraction of first neighbors.

**Performance**

[Fig. 8.] shows averaged performance increase due to the same simulation performed on GPU and CPU.

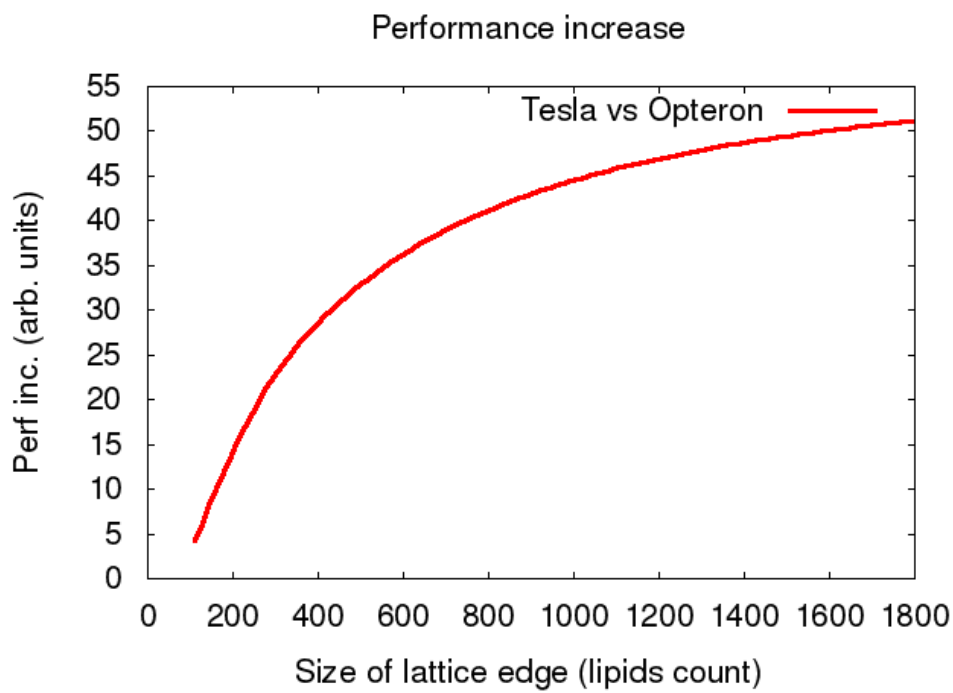

Fig. 8. Performance increase CPU vs GPU computations.

**Cluster Analysis**

Average cluster size is the time function of size of the system for fixed omega and lipid composition. As our research shows, there is a need of performing simulations on larger domains, even for very simple models, as cluster sizes can be significantly big for 400x400 lattices. Thus, the cluster size distribution can suffer from artifacts due to the lattice size [Fig. 9.] The average cluster size distribution also varies for smaller lattice sizes [Fig. 10].

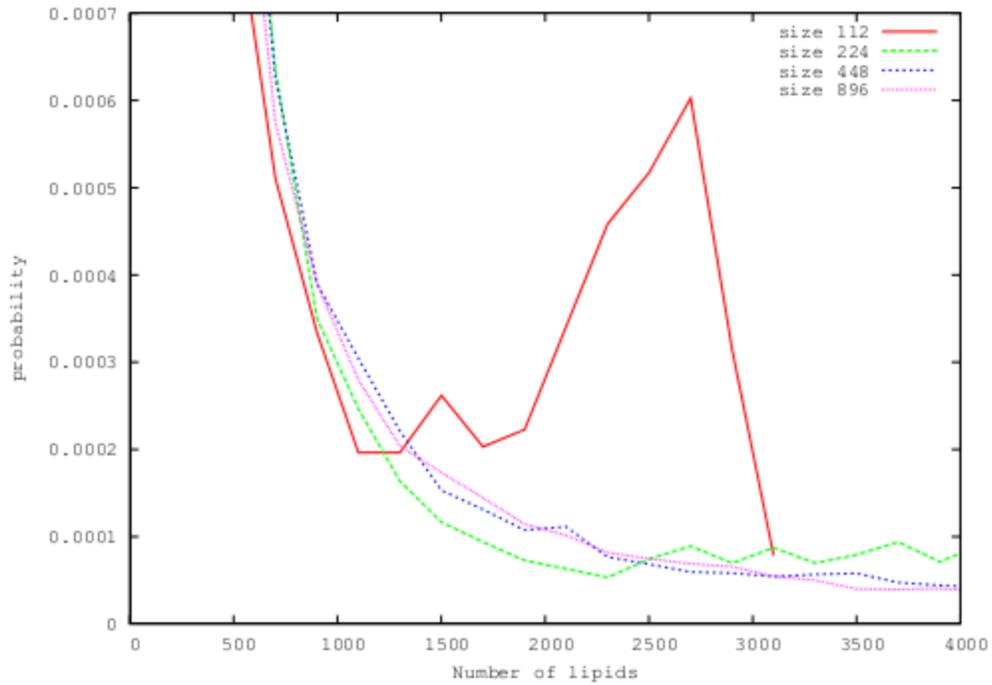

Fig. 9. Artifacts due to small size of simulation lattice presented on cluster size distribution plot.

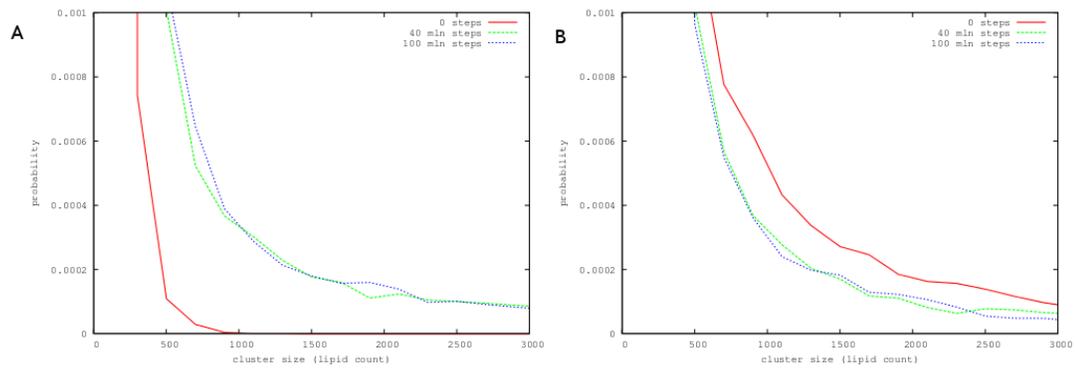

Fig. 10. Comparison in cluster sizes distribution between random and non-random start of simulation.

More complex system simulation (e.g. including cholesterol) will probably need even bigger lattice size to avoid such artifacts.

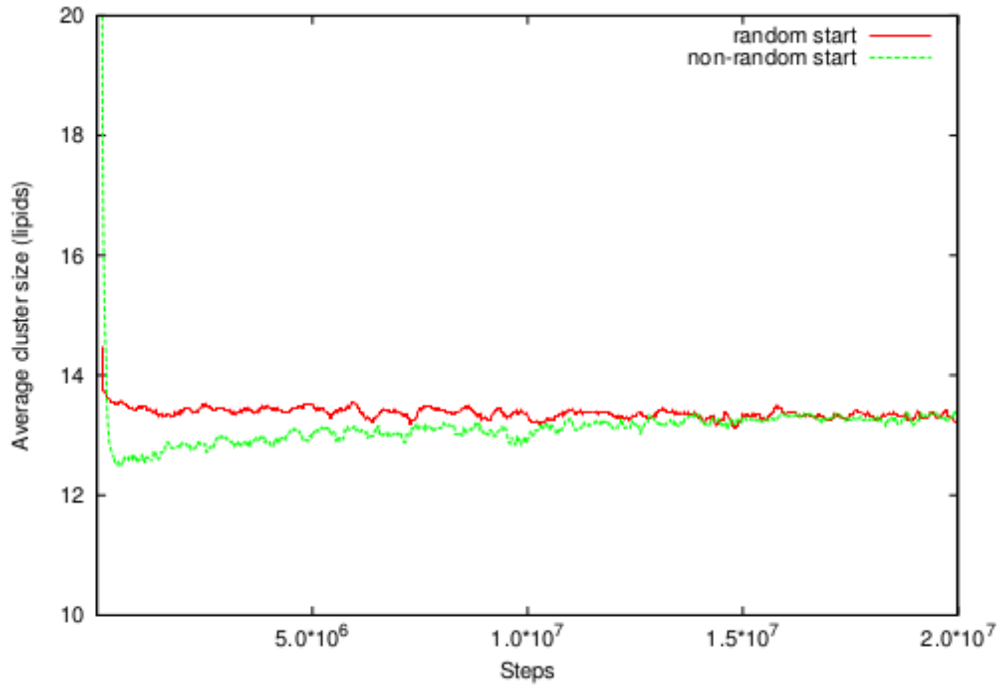

Fig. 11. Average cluster size comparison in simulation from random and non-random started simulation

As results show, the average cluster size is a stable and well defined characteristic, as it does not depend on starting condition [Fig. 11.].

As the results proved (not shown) for larger omegas that it is even more important to perform simulations on larger size of the lattice as the artifact tend to increase with the omega parameter. This can also be reflected to model complexity level - the more elements are taken into consideration, the larger simulation lattice is required. Especially when it comes to comparison with experiment, where Giant Unilamellar Vesicles are usually involved, the size of the lattice plays an important role.

**Fraction of first neighbors calculated on images**

Fig. 12 presents comparison between results of calculation of fraction of first neighbors using raw data and using images generated with an algorithm described in previous sections. Figure shows that results are acceptable, in particular, the difference between results of calculations differs no more than 0.1%.

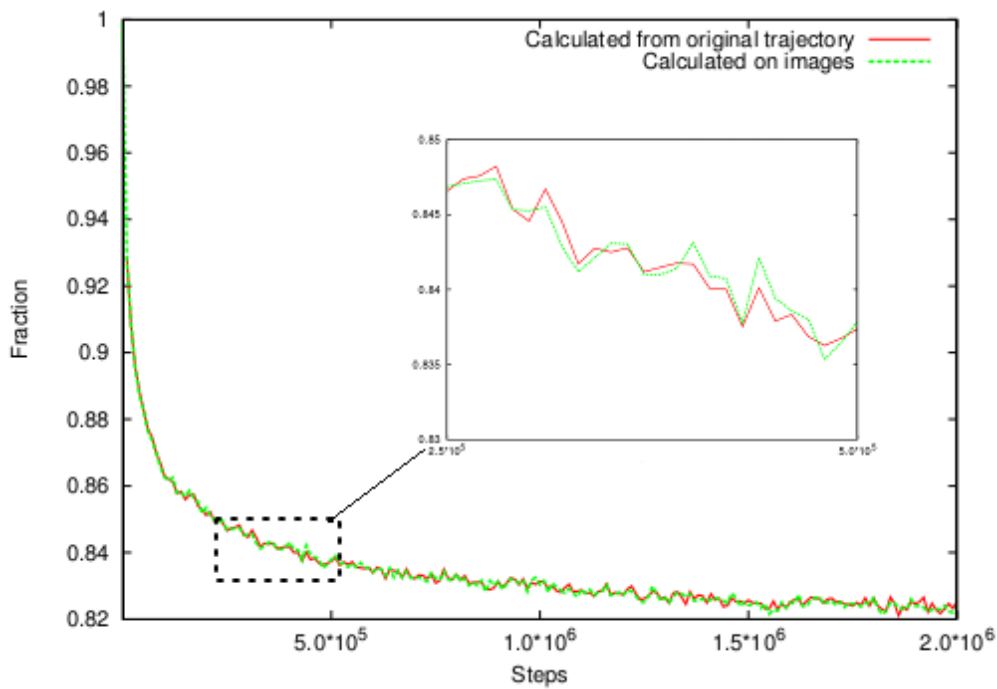

Fig. 12. Comparison between fraction of first neighbors calculated from images and from raw data on a non-random start trajectory.

## Discussion

In this article, we show effective algorithm for computational studies utilizing Kawasaki Kinetics. GPU - based implementation appears to be ~55 times faster than sequential while preserving the same result characteristics. Comparison between CPU based and GPU implementation is provided to show correctness of proposed algorithm. Simple Monte Carlo model for lipid microdomain formation is provided for showing analysis methods. Time-dependent analysis of cluster formation in lipid systems is provided.

To simulate larger systems with rafts and proteins, there is a need for larger simulation sizes thus proposed algorithm can be useful.